\documentclass[fleqn,12pt,twoside]{article}
\usepackage[headings]{espcrc1}

\readRCS
$Id: espcrc1.tex,v 1.2 2004/02/24 11:22:11 spepping Exp $
\usepackage{graphicx}
\usepackage[figuresright]{rotating}

\newcommand{\AmS}{{\protect\the\textfont2
  A\kern-.1667em\lower.5ex\hbox{M}\kern-.125emS}}

\newcommand{\lsim}{\,{\buildrel < \over {_\sim}}\,}
\newcommand{\gsim}{\,{\buildrel > \over {_\sim}}\,}
\newcommand{\sqrtsNN}{\sqrt{s_{\scriptscriptstyle{{\rm NN}}}}}
\newcommand{\av}[1]{\left\langle #1 \right\rangle}

\newcommand{\mev}{\mathrm{MeV}}
\newcommand{\gev}{\mathrm{GeV}}

\newcommand{\cm}{\mathrm{cm}}

\newcommand{\PbPb}{\mbox{Pb--Pb}}

\newcommand{\RCP}{R_{\rm CP}}
\newcommand{\pt}{p_T}

\renewcommand{\d}{{\rm d}}
\newcommand{\dNdy}{{\rm d}N_{\rm ch}/{\rm d}y}

\newcommand{\Kzs}{{\rm K^0_S}}
\newcommand{\La}{\Lambda}
\newcommand{\Al}{\overline\Lambda}
\newcommand{\decayarrow}{\makebox[0mm][l]{\rule{0.33em}{0mm}\rule[0.55ex]{0.044em}{1.55ex}}\rightarrow}

\let\Otemize =\itemize
\let\Onumerate =\enumerate
\let\Oescription =\description
\def\Nospacing{\itemsep=0pt\topsep=0pt\partopsep=0pt\parskip=0pt\parsep=0pt}
\def\Topspac{\vspace{-0.5\baselineskip}}
\def\Botspac{\vspace{-0.2\baselineskip}}
\newenvironment{Itemize}{\Topspac\Otemize\Nospacing}{\endlist\Botspac}

\title{Results from NA57}

\author{Andrea Dainese
          \address{Dipartimento di 
              Fisica ``G.~Galilei'', Universit\`a
                 degli Studi di Padova, Padova, Italy}
         ~(for the NA57 Collaboration)\thanks{For the full NA57 
                   Collaboration author list, see
                appendix `Collaborations' of this volume.}}
       
\runtitle{Results from NA57}
\runauthor{A.~Dainese, for the NA57 Collaboration}

\begin{document}

\maketitle

\begin{abstract}
The NA57 experiment has measured the production of strange and multi-strange 
hadrons in heavy-ion collisions at the CERN SPS. After briefly introducing 
the NA57 apparatus and analysis procedures, we present recent results 
on strangeness enhancement in \mbox{Pb--Pb} relative to \mbox{p--Be} 
collisions, 
on the study of the $m_T$ distributions of strange particles,
and on central-to-peripheral nuclear modification factors
in \mbox{Pb--Pb} collisions at top SPS energy. 
\end{abstract}

\section{Introduction}
\label{intro}

The measurement of strangeness production in nucleus--nucleus collisions 
at high energy is a valuable tool to study the properties of the 
dense system expected to be formed in the collision.
 
The enhancement of the global
production yield of strange and multi-strange baryons in nucleus--nucleus
relative to proton-induced reactions was indicated,
already at the beginning of the Eighties, 
 as a signature 
for the phase transition from a hadron gas to a deconfined state of
quarks and gluons (the quark-gluon plasma)~\cite{rafelski}. 
Enhancements increasing with the strangeness content
of the particle were first observed by WA97
at $158~A~\gev/c$ beam momentum~\cite{wa97v0sele}.
The NA57 results, that we will present in section~\ref{enh},
extend those of WA97 to a wider centrality range and
to lower beam momentum of $40~A~\gev/c$.

The transverse mass, $m_T=\sqrt{m^2+p_T^2}$, and rapidity 
distributions of particles produced in nucleus--nucleus collisions 
are expected to be sensitive to the transverse and 
longitudinal expansion dynamics of the system formed in the 
collision~\cite{BlastRef}.
Information on these dynamical effects can 
be extracted from the analysis of the shapes of the bulk 
kinematic distributions of strange particles (section~\ref{dyn}).
  
At large transverse momenta (for SPS energy) of $2$--$4~\gev/c$ 
the comparison of the binary-scaled $\pt$-differential yields of strange 
particles in central and in peripheral collisions 
may provide information on the mechanisms at play 
in the in-medium propagation and hadronization of energetic strange quarks
(section~\ref{rcp}).

\section{Experimental setup and data analysis}
\label{exp}

The NA57 apparatus~\cite{na57setup} was
designed to study the production of strange and multi-strange
hadrons in fixed-target heavy-ion collisions by reconstructing
their weak decays into final states containing charged particles only:
\[
\begin{array}{lllllllllllllll}
    \Kzs &\rightarrow & \pi^+ + \pi^-  & \hspace{1mm}&
    \La  &\rightarrow & {\rm p} + \pi^- & \hspace{1mm}&
    \Xi^- &\rightarrow & \La + \pi^- &  \hspace{1mm}&  \Omega^- &\rightarrow &
   \La + {\rm K}^-  \\
         &&&&&&&   &        &    &
   \decayarrow  {\rm p} + \pi^-  &  &  &  & \decayarrow  {\rm p} + \pi^- 
\end{array}
\]
(and charge-conjugates, for the hyperons).
Tracks are reconstructed in the $5\times 5\times 30~\cm^3$ 
silicon pixel detector telescope
which is placed 60~cm downstream of the target.
Its acceptance covers about half a unit in
rapidity at mid-rapidity and transverse momentum larger than
about $0.5~\gev/c$. 
The centrality trigger,
based on charged multiplicity,
was set so as to select approximately the most central 60\% of the inelastic
collisions. 

The strange particles selection procedure is described in detail
in~\cite{wa97v0sele,mtpaper}. The main decay-vertex identification criteria 
are the following:
(a) the two decay trajectories 
    are compatible with the hypothesis of having a common
    origin point;
(b) the reconstructed decay vertex is well separated from the
      target.
NA57 does not have charged-particle identification, but 
ambiguities among $\Kzs$, $\La$ and $\Al$ are completely eliminated 
by means of kinematic cuts~\cite{wa97v0sele}.
The high track-parameters resolution provided by the silicon pixel detectors
allows to obtain final signal samples that are essentially background-free
(see~\cite{mtpaper} and~\cite{rcppaper}).
Negatively charged particles, $h^-$, for the analysis 
presented in section~\ref{rcp} are selected requiring that they
point back to the interaction vertex, with a 
residual contamination of secondary tracks 
estimated to be smaller than 2\%.
For the measurement of the production yields and of the 
kinematic distributions (sections~\ref{enh} and~\ref{dyn})
we corrected for acceptance and efficiency by assigning 
to every selected particle a weight,
calculated on the basis of a Monte Carlo simulation~\cite{mtpaper}.

The collision centrality is determined using the charged particle
multiplicity $N_{\rm ch}$ in the pseudorapidity range $2<\eta<4$,
sampled by the microstrip silicon detectors (MSD) 
as described in~\cite{wounded,mult2004}.
We fit the $N_{\rm ch}$-differential \mbox{Pb--Pb} cross section
$\d\sigma/\d N_{\rm ch}$, as selected by the centrality trigger, assuming
$N_{\rm ch}=q\cdot N_{\rm part}^\alpha$ 
(a modified Wounded Nucleon model)~\cite{wounded}, where
$N_{\rm part}$ is the number of participants, i.e.\,nucleons participating
in the primary nucleon--nucleon collisions, 
estimated from the Glauber model. 
The sample of collected events is subdivided in centrality classes, 
with $N_{\rm ch}$ limits corresponding to given
fractions of the \mbox{Pb--Pb} inelastic cross section. For each class the average
number of participants, $\av{N_{\rm part}}$, and of binary 
collisions, $\av{N_{\rm coll}}$ are calculated, after the fit, 
from the Glauber model.
For \mbox{p--Be} and \mbox{p--Pb} collisions, the values of $\av{N_{\rm part}}$ 
are calculated directly from the Glauber model. 



\section{Enhancements: How is strangeness cooked in the system?}
\label{enh}

For each particle species, the corrected double-differential distribution
is fitted to the expression: 
\begin{equation}
\label{eq:expo}
\d^2N/\d m_{T} \d y=f(y)\,m_{T}\,\exp\left(-m_{T}/T_{\rm app}\right)\,,
\end{equation}
where, within our acceptance of about a unit in rapidity around mid-rapidity
$y_{\rm cm}$, 
the shape of the rapidity distribution $f(y)$ is taken to be gaussian 
for $\Kzs$ and $\Al$, and flat for the other strange 
particles~\cite{rapiditypaper}. The yield $Y$ is calculated by integrating
the fit expression over the phase-space region 
$\{0<\pt<\infty\}\times \{y_{\rm cm}-0.5<y<y_{\rm cm}+0.5\}$. 
The enhancement $E$ is then defined as the yield per participant in a given 
centrality class in \mbox{Pb--Pb} divided by the yield per participant in \mbox{p--Be}:  
\begin{equation}
E={\left(  Y/\av{N_{\rm part}}  \right)_{\rm Pb-Pb}} \Big/ {
   \left(  Y/\av{N_{\rm part}}  \right)_{\rm p-Be}     }
\label{eq:enh} 
\end{equation} 
For the strangeness enhancement analysis we considered five centrality classes
corresponding to the following intervals in percentiles of the 
\mbox{Pb--Pb} inelastic cross section, 
from the most central to the most peripheral: 
0--4.5\%, 4.5--11\%, 11--23\%, 23--40\%, 40--53\%.
The enhancements in \mbox{p--Pb} relative to \mbox{p--Be} collisions 
are computed as well.

\begin{figure}[!t]
\centering
\includegraphics[width=.7\textwidth]{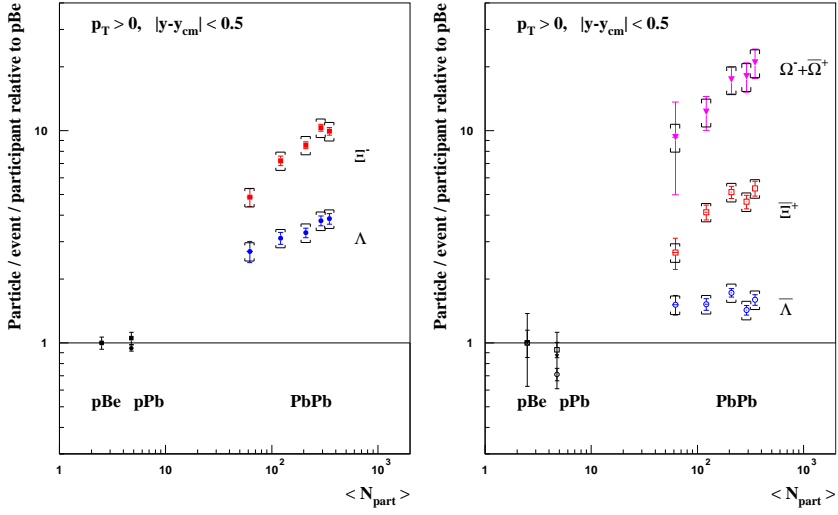}
\caption{\label{fig:enhancements160} Centrality dependence of hyperon
enhancements at 158 $A$ GeV/$c$. 
The bars indicate the statistical errors,
while the bracket symbols represent the systematic errors.} 
\end{figure}

In Fig.~\ref{fig:enhancements160} the hyperon 
enhancements at 158~$A$~GeV/$c$ are shown as a function of $\av{N_{\rm part}}$.
No enhancement in the yield per participant is observed when going from 
\mbox{p--Be} to \mbox{p--Pb}, while  
a clear pattern of increasing enhancement with increasing strangeness
content is observed in \mbox{Pb--Pb},
up to an enhancement of a factor about 20 
for the triply-strange $\Omega$ in the most central collisions.
We note that the enhancements increase with centrality for all hyperons, 
except $\Al$, which is even suppressed in \mbox{p--Pb} relative to \mbox{p--Be}.

The pattern $E(\La)<E(\Xi)<E(\Omega)$ is in agreement with the 
`historic' predictions for a QGP scenario~\cite{rafelski}, the rationale 
behind these predictions being that the $s$ and $\overline s$ quarks, 
abundantly produced in the deconfined phase would recombine to 
form strange and multi-strange particles in a time much shorter than 
that required to produce them by successive rescattering interactions in a
hadronic gas.

\begin{figure}[!t]
\centering
\includegraphics[width=0.32\textwidth]{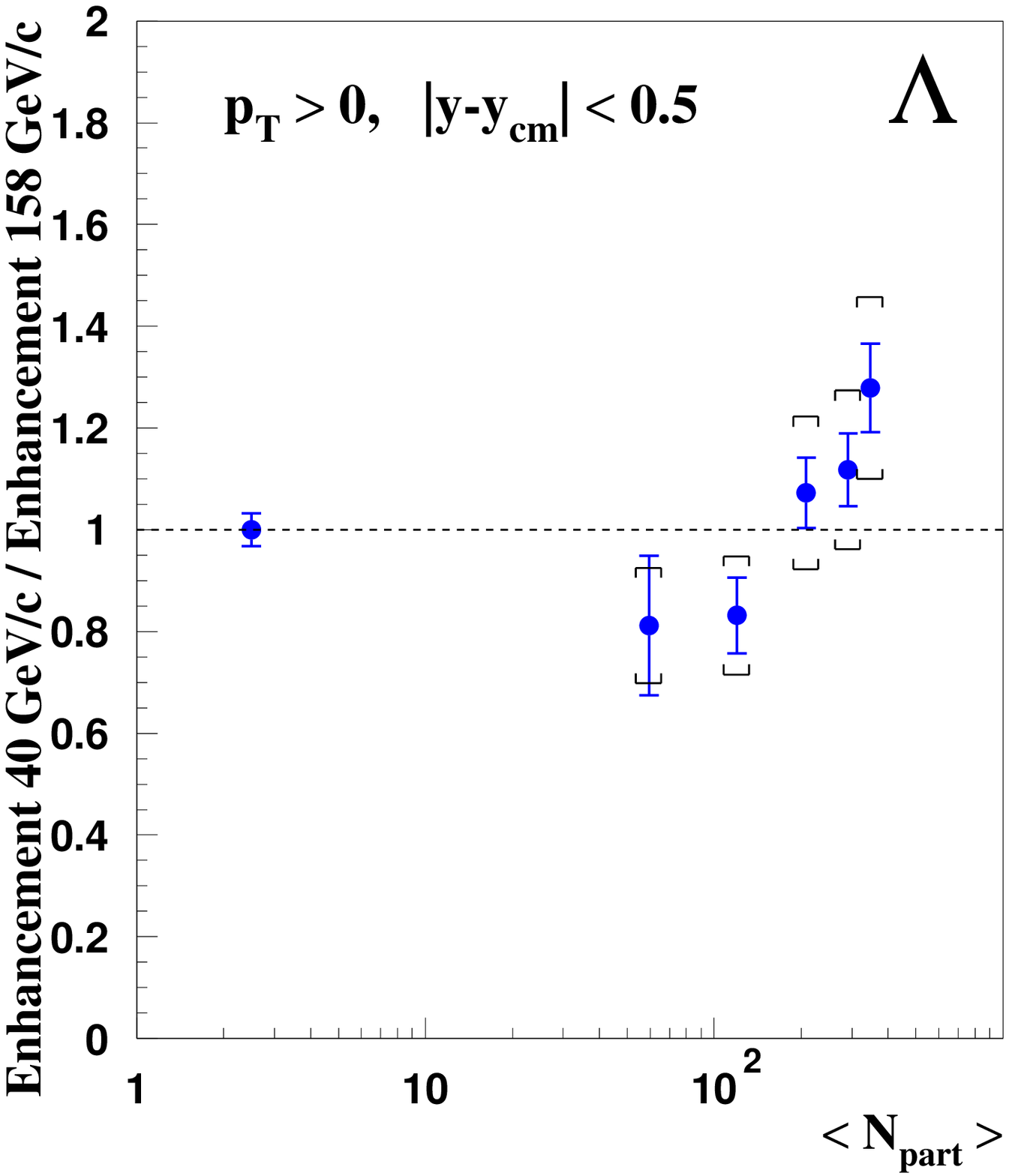}
\includegraphics[width=0.32\textwidth]{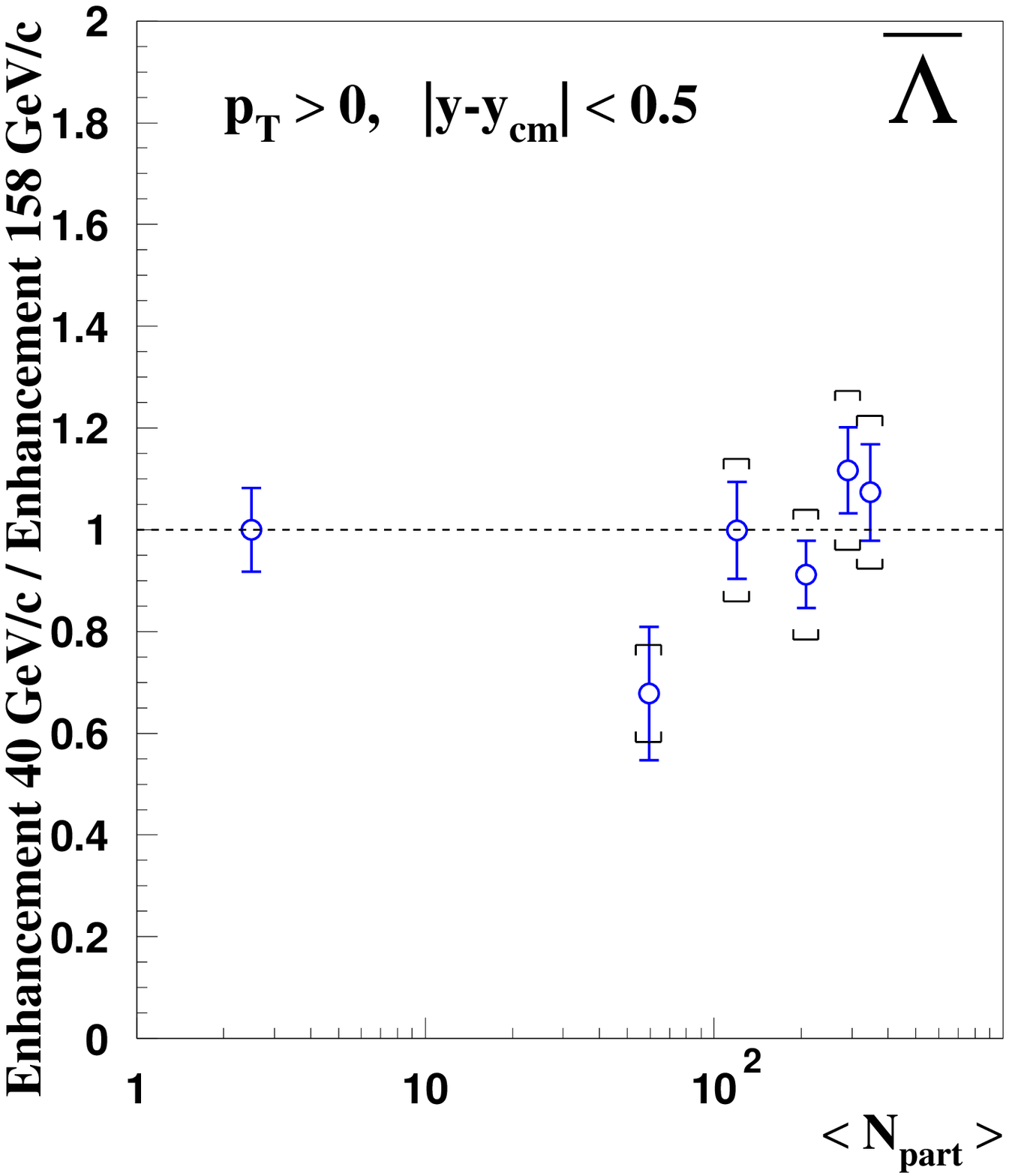}
\includegraphics[width=0.32\textwidth]{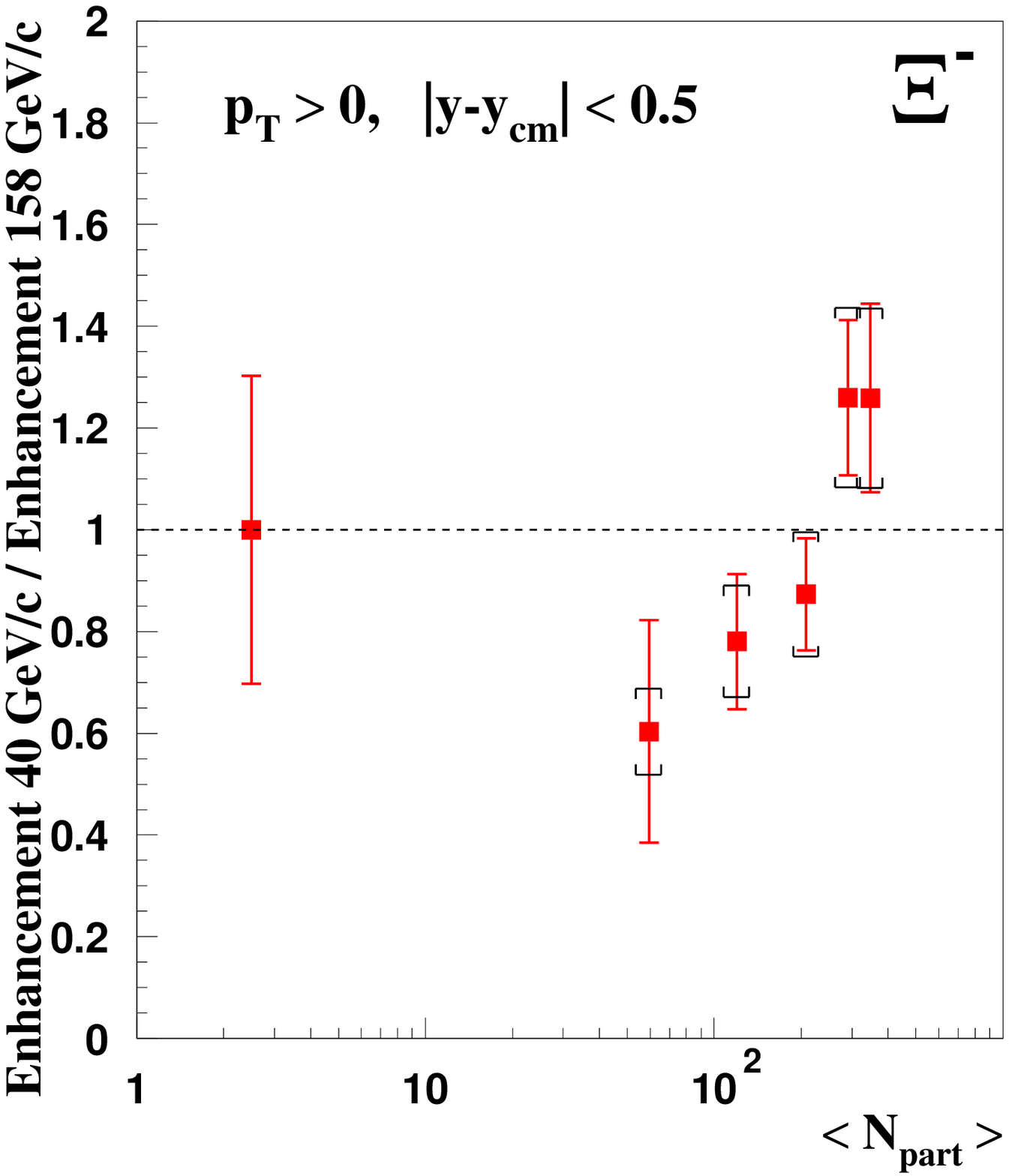}
\caption{\label{fig:enhRatios} Ratios of the enhancements 
at 40 $A$~GeV/$c$ to those at 158 $A$~GeV/$c$, as 
a function of centrality, for $\La$, $\Al$, and $\Xi^-$
(error symbols as in Fig.~\ref{fig:enhancements160}).} 
\end{figure}

We illustrate the energy-dependence of the enhancements by showing the
ratio of the enhancements at 40 $A$~GeV/$c$ to those at 158~$A$ GeV/$c$, 
in Fig.~\ref{fig:enhRatios}, for $\La$, $\Al$ and $\Xi^-$ 
(due to the limited statistics in \mbox{p--Be} collisions, the 40~$A$~GeV/$c$ 
enhancements
could not be calculated for rarer particles, $\overline \Xi^+$ and $\Omega$).
The enhancement is similar at the two energies, although 
its increase with centrality is steeper at 40~$A$~GeV/$c$, where 
 we have indications, for the $\Xi^-$ in particular, 
for a larger enhancement than at 158~$A$~GeV/$c$
in the two most central classes.
This experimental observation is in agreement with the prediction of
a theoretical model~\cite{Redlich}, where the strangeness enhancement is 
obtained as a consequence of the removal of canonical suppression, 
when going from p--A to \mbox{Pb--Pb} collisions
(see, e.g.,~\cite{federico} for a discussion).
Note, however, that the model~\cite{Redlich} does not reproduce correctly 
the centrality dependence 
nor the absolute magnitude of the enhancements.

\section{$m_T$ and $y$ spectra: How does the system expansion 
affect strange particles?}
\label{dyn}

The bulk of the kinematical 
distributions for particles produced in nucleus--nucleus
interactions
is expected to be shaped by the superposition of 
two effects: the thermal motion of the particles in the {\it fireball} 
and a pressure-driven radial (for $m_T$) or longitudinal (for $y$) 
collective flow, induced by the fireball expansion.  
Studies of the  
$m_T$ spectra for $\La$, $\Xi$, $\Omega$ hyperons, 
 and $\Kzs$, measured by NA57 in \mbox{Pb--Pb} collisions at 
158 and 40 $A$\ GeV/$c$ were presented in~\cite{mtpaper} 
and~\cite{hq04giu}, respectively. A study of the rapidity distributions
in the NA57 acceptance, $|y-y_{\rm cm}|<0.5$, at 158~$A$~$\gev/c$
was presented in~\cite{rapiditypaper}.

\begin{figure}[!t]
\centering
\resizebox{1.00\textwidth}{!}{%
\includegraphics{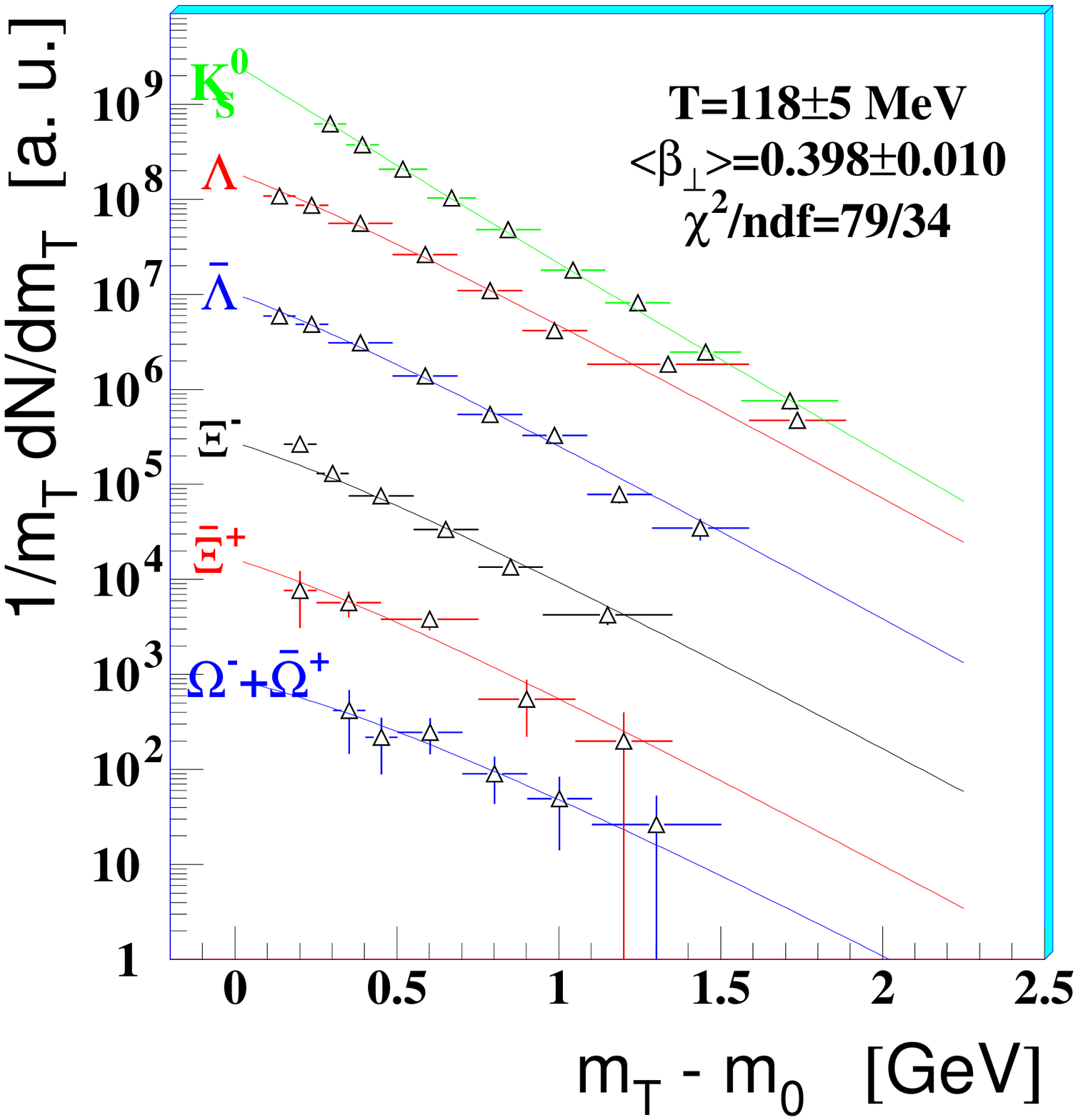}
\includegraphics{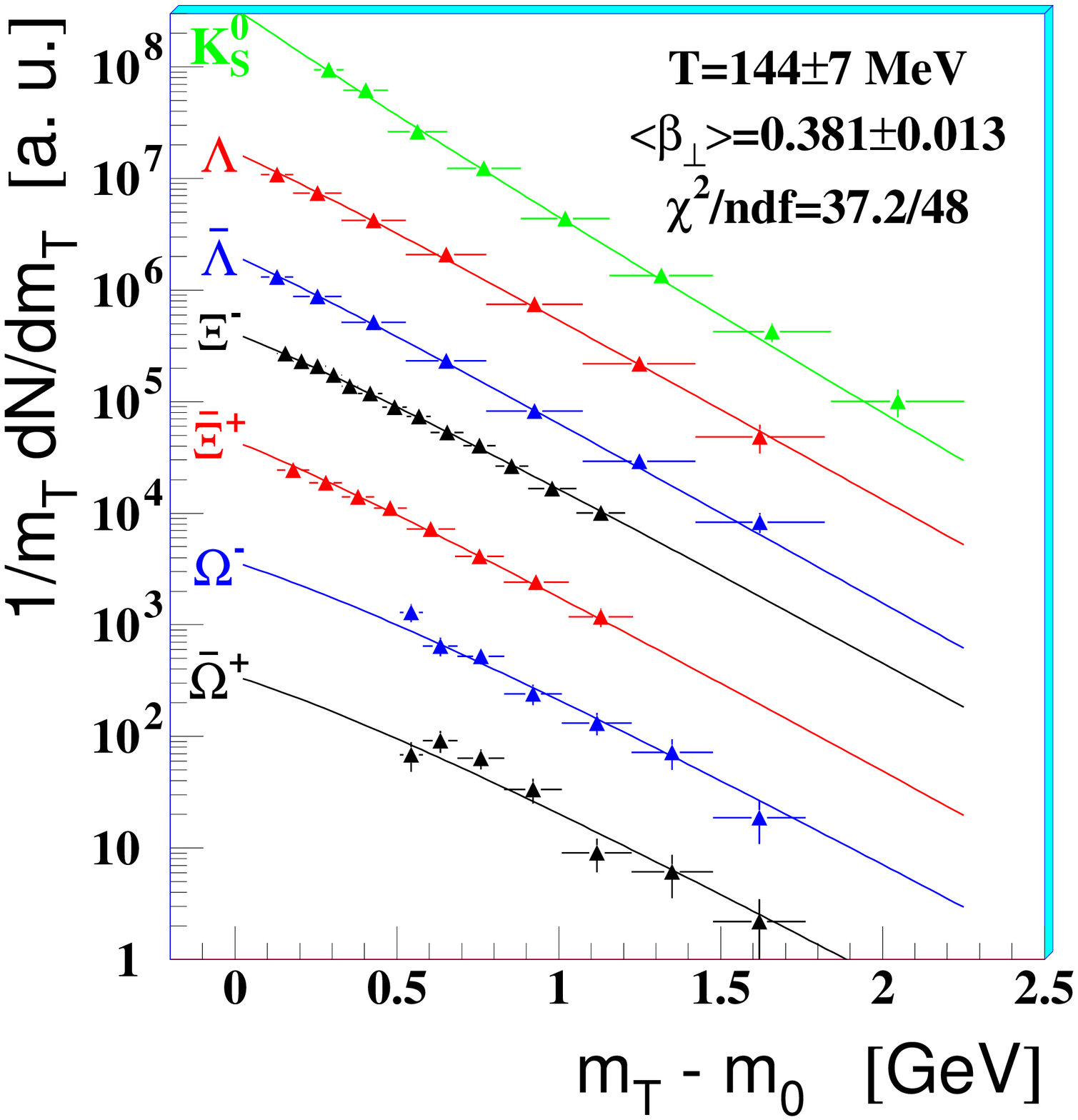}
\includegraphics{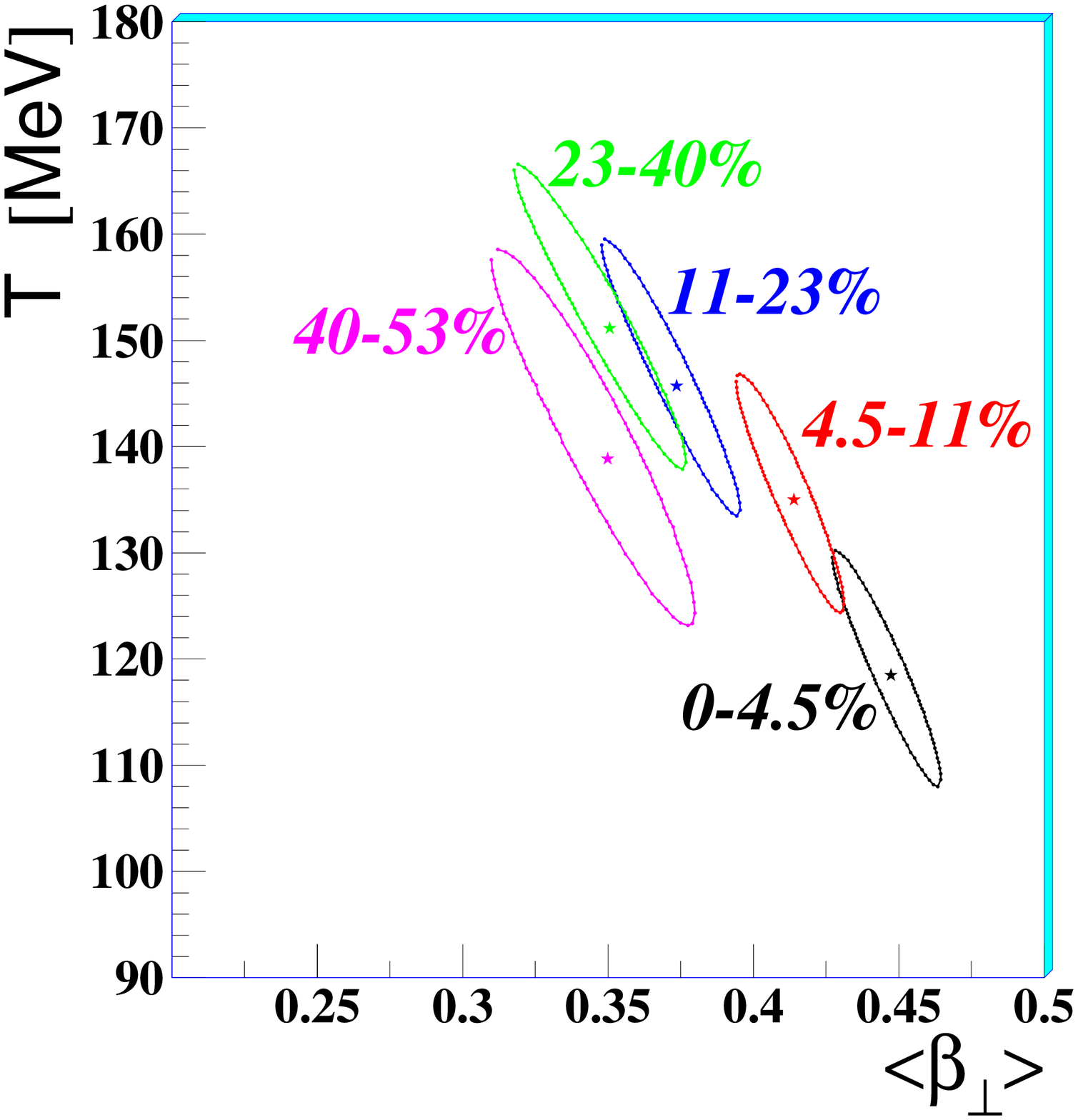}
}
\caption{Blast-wave fits to the 
          transverse mass spectra of strange particles for the 53\%
          most central \mbox{Pb--Pb} collisions at 158 (left) and at 40  
	  (middle) $A$\ GeV/$c$. 
	  Right: centrality dependence of $T$ and $\av{\beta_{\perp}}$
          at 158~$A$~$\gev/c$ (1\,$\sigma$ contour 
          plots)~\cite{mtpaper,hq04giu}.}
\label{fig:blast}
\end{figure}

The $m_T$ spectra were analyzed in the framework of the {\it blast-wave} 
model~\cite{BlastRef}. The model 
assumes cylindrical symmetry for an expanding fireball in local  
thermal equilibrium and predicts the shape of the 
double-differential yield $\d^2N/\d m_T \d y$ 
for the different particle species, in terms of the kinetic freeze-out 
temperature $T$ and of  
the radial velocity profile $\beta_{\perp}(r)$, 
that we parametrized as $\beta_{\perp}(r) = \beta_S\cdot r/R_G$,
being $\beta_S$ the flow velocity at the surface and $R_G$ the 
outer radius.  
The fit to 
the experimental spectra allows to extract $T$ and the average
transverse flow velocity $\av{\beta_{\perp}}=2/3\cdot\beta_S$~\cite{mtpaper}. 
The results of the simultaneous fits to all particle species, 
centrality-integrated over the range 0--53\% covered by NA57,  
are shown in Fig.~\ref{fig:blast} for collisions at 40 (left) and 
158~$A$~$\gev/c$ (centre). The resulting kinetic freeze-out temperature 
and average transverse flow velocity values are:
\[
T=(118\pm5\pm11)~\mev\,,~~\av{\beta_\perp}=0.38\pm0.01\pm0.01~~~~{\rm at}~40~A~\gev/c
\]
\[
T=(144\pm7\pm14)~\mev\,,~~\av{\beta_\perp}=0.40\pm0.01\pm0.01~~~~{\rm at}~158~A~\gev/c
\]
where the first error is statistical and the second is systematic.
A lower thermal freeze-out temperature is measured at lower collision energy, 
while  
the transverse flow velocities are found to be compatible within errors.  
When performed separately in the five centrality classes at 158~$A~\gev/c$,
the analysis shows an increase of $T$ and a decrease of $\av{\beta_\perp}$ 
from central (0--4.5\%) to 
semi-peripheral (40--53\%) collisions, as presented in the 
Fig.~\ref{fig:blast} (right).
This may be interpreted as an indication for earlier freeze-out of the 
system at lower energy.

In order to extract information about the longitudinal expansion dynamics 
from the rapidity distributions, we used 
the blast-wave model~\cite{BlastRef}, 
with a Bjorken-expansion scenario~\cite{Bjorken} folded with a thermal  
distribution of the fluid elements (see~\cite{rapiditypaper} for details
of the analysis). Since, within the limited rapidity acceptance
of the NA57 telescope ($|y-y_{\cm}|<0.5$),
the sensitivity is not sufficient to  
constrain both the kinetic freeze-out temperature and 
the longitudinal flow velocity, we fixed the temperature to the 
value of about $144~\mev$ extracted from the $m_T$ analysis, and we 
fitted only the longitudinal flow velocity.
The distributions of all strange particles under study 
can be fitted simultaneously 
($\chi^2/{\rm ndf} \approx 1$)~\cite{rapiditypaper}.
The resulting average longitudinal flow velocity is  
$\av{\beta_{\rm L}}=0.42\pm0.03(stat)$, 
similar to $\av{\beta_\perp}$ (see above), suggesting large  
nuclear stopping along the beam direction.


\section{$\RCP$: How does the medium affect $s$-quarks 
         propagation and hadronization?}
\label{rcp}

At RHIC energies, 
the central-to-peripheral nuclear modification factor
\begin{equation}
\RCP(\pt) = {\av{N_{\rm coll}}_{\rm P} \over \av{N_{\rm coll}}_{\rm C}}\times
\frac{\d^2 N_{\rm AA}^{\rm C}/\d\pt\d y}{\d^2 N_{\rm AA}^{\rm P}/\d\pt\d y}
\label{eq:rcp}
\end{equation}
has proven to be a powerful tool for the study of parton propagation 
in the dense system formed in nucleus--nucleus collisions
 (see, e.g.,~\cite{starRcpk0la}).
At high $p_T$ ($\gsim 7~\gev/c$), 
$\RCP$ is found to be suppressed by a factor 3--4 with respect 
to unity (at $\sqrtsNN=200~\gev$), for all particle species;
this is interpreted as a consequence of parton energy loss in the medium, 
prior to fragmentation {\it outside} the medium.
At intermediate transverse momenta ($2$--$5~\gev/c$), instead, partons
are believed to lose energy {\it and} 
hadronize {\it inside} the medium via the mechanism of recombination; this
would originate the measured pattern of smaller suppression for 
baryons relative to mesons. 

The study of $\RCP$ and of its particle-species dependence 
at top SPS energy allows to test for these 
phenomena at an energy smaller by about one order of magnitude  
($\sqrtsNN=17.3~\gev$). While measurements of the $\pi^0$ $\RCP$
by the WA98 Collaboration, supporting the presence 
of parton energy loss effects, were
 published already in 2002~\cite{wa98}, the first data on the 
particle-species dependence have been 
presented by the NA57 Collaboration in~\cite{rcppaper} 
and by the NA49 Collaboration 
at this Quark Matter 2005 conference~\cite{hoene}. 

\begin{figure}[!t]
\begin{center}
   \includegraphics[width=.49\textwidth]{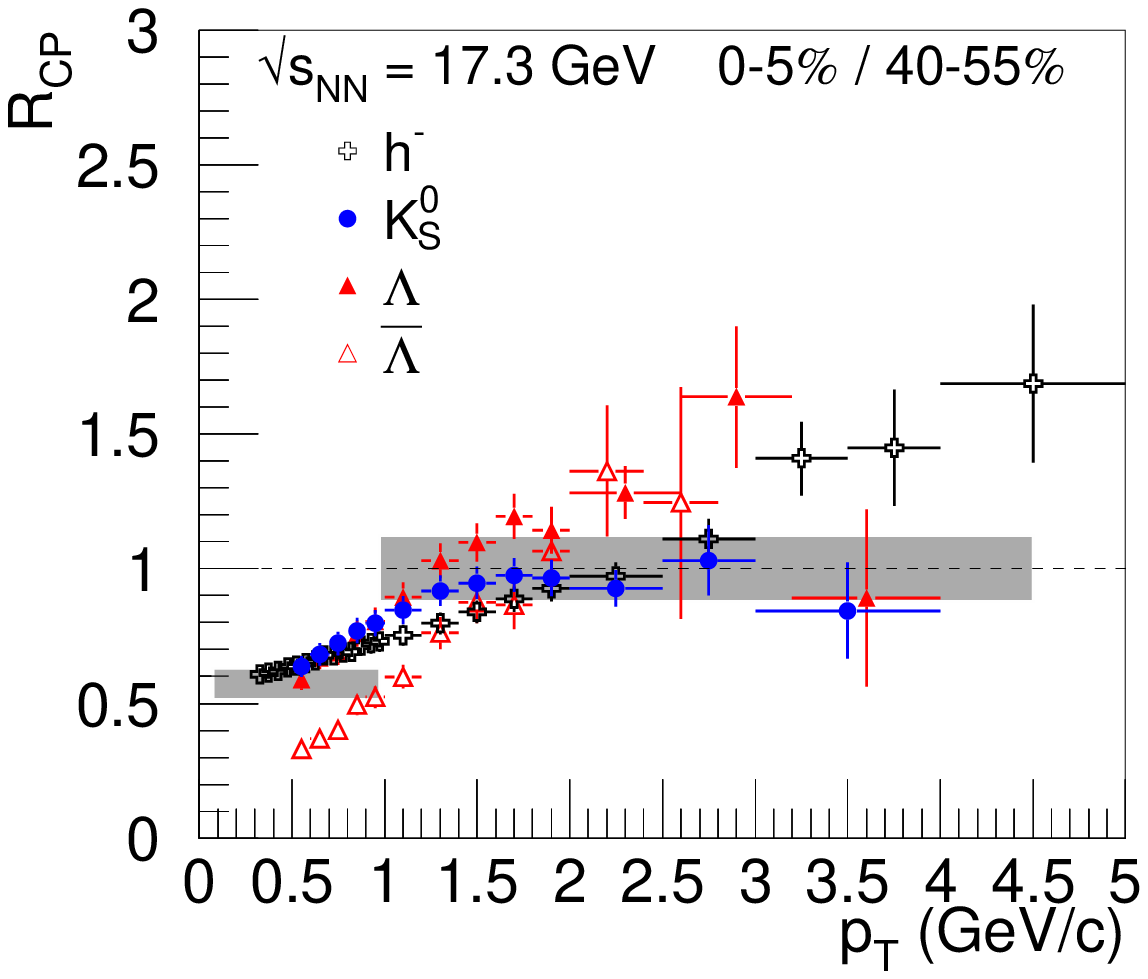}
   \includegraphics[width=.49\textwidth]{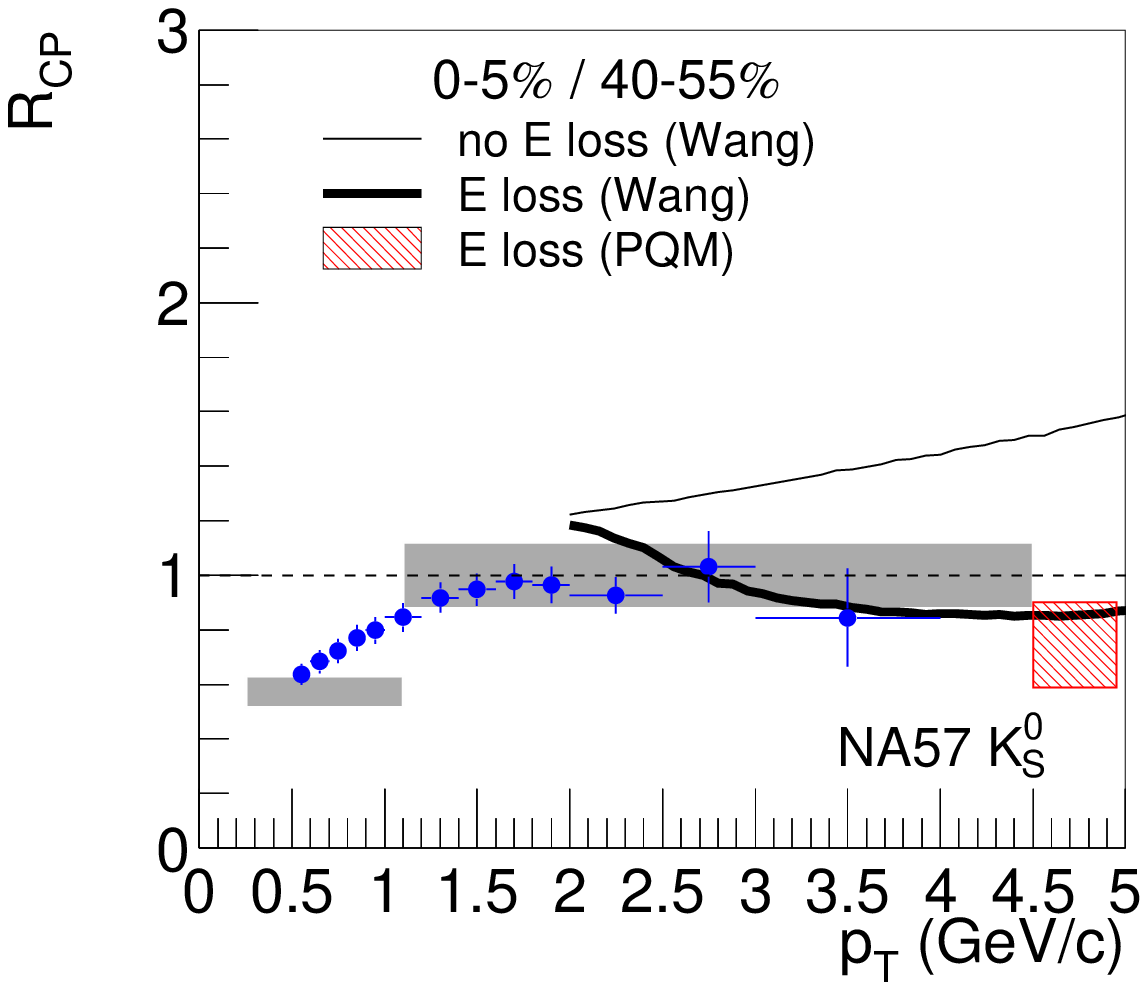}
   \caption{Left: $\RCP(\pt)$
            for $h^-$, $\Kzs$, $\La$ and $\Al$
            in \PbPb~collisions at $\sqrtsNN=17.3~\gev$~\cite{rcppaper};
            bars are the sum of statistical and 
              point-by-point systematic errors;
            shaded bands centered
            at $\RCP=1$ represent the systematic error due to the uncertainty 
            in the ratio of the
            values of $\av{N_{\rm coll}}$ in each class; shaded
            bands at low $\pt$ represent the values expected for
            scaling with the number of participants, with their 
            systematic error. Right: the $\Kzs$ $\RCP(\pt)$
            compared to predictions~\cite{wang,pqm}
            with and without energy loss.}
   \label{fig:Rcp}
\end{center}
\end{figure}

In order to exploit the full sample of collected data,
we calculate $\RCP(\pt)$ for $h^-$, $\Kzs$, $\La$ and $\Al$ using $\pt$
distributions which are not corrected for geometrical acceptance 
and reconstruction/selection efficiency.
We have verified~\cite{rcppaper} on a subsample of events 
that the corrections
do not depend on the event
centrality over the full transverse momentum range covered, 
$0.5<\pt\lsim 4~\gev/c$.
Figure~\ref{fig:Rcp} (left) shows the results for the 0--5\%/40--55\%
$\RCP$. The results are in qualitative agreement with those presented
by NA49~\cite{hoene} for charged pions, charged kaons and protons plus
antiprotons (a quantitative comparison of NA57 $\Kzs$ and NA49 $\rm K^\pm$
is at the moment not straightforward, since different definitions of the 
reference peripheral class are used by the two experiments).

In the right-hand panel of Fig.~\ref{fig:Rcp} 
we compare our $\Kzs$ data to predictions provided by
X.N.~Wang, obtained from a perturbative-QCD-based
calculation~\cite{wang}, including (thick line) or
excluding (thin line) in-medium parton energy loss. The initial
gluonic rapidity density of the medium, $\d N_{\rm g}/\d y$, 
was scaled down from that
needed to describe RHIC data, according to the decrease by
about a factor 2 in $\dNdy$ from 200 to 17.3~GeV c.m.s. energy.
The curve without energy loss shows a
large `Cronin-enhancement' toward high $\pt$,
included in the calculation via an initial-state partonic intrinsic
transverse momentum broadening, tuned on the 
original Cronin effect data~\cite{cronin}.
This enhancement is not present in our $\Kzs$ data,
that are better described by the curve including energy loss. 
As a cross-check, we compared the $\RCP$
value predicted by X.N.~Wang with energy loss to the prediction of an
independent model of parton energy loss, the Parton Quenching
Model (PQM),
that describes several energy-loss-related observables at RHIC
energies~\cite{pqm}.
The two models predict a similar energy loss
effect at SPS energy, i.e.\,a reduction  of
the 0--5\%/40--55\% $\RCP$ (for $\pt\gsim 4~\gev/c$)
by about a factor 2, with respect to the value calculated without 
energy loss.

\begin{figure}[!t]
\begin{center}
   \includegraphics[width=\textwidth]{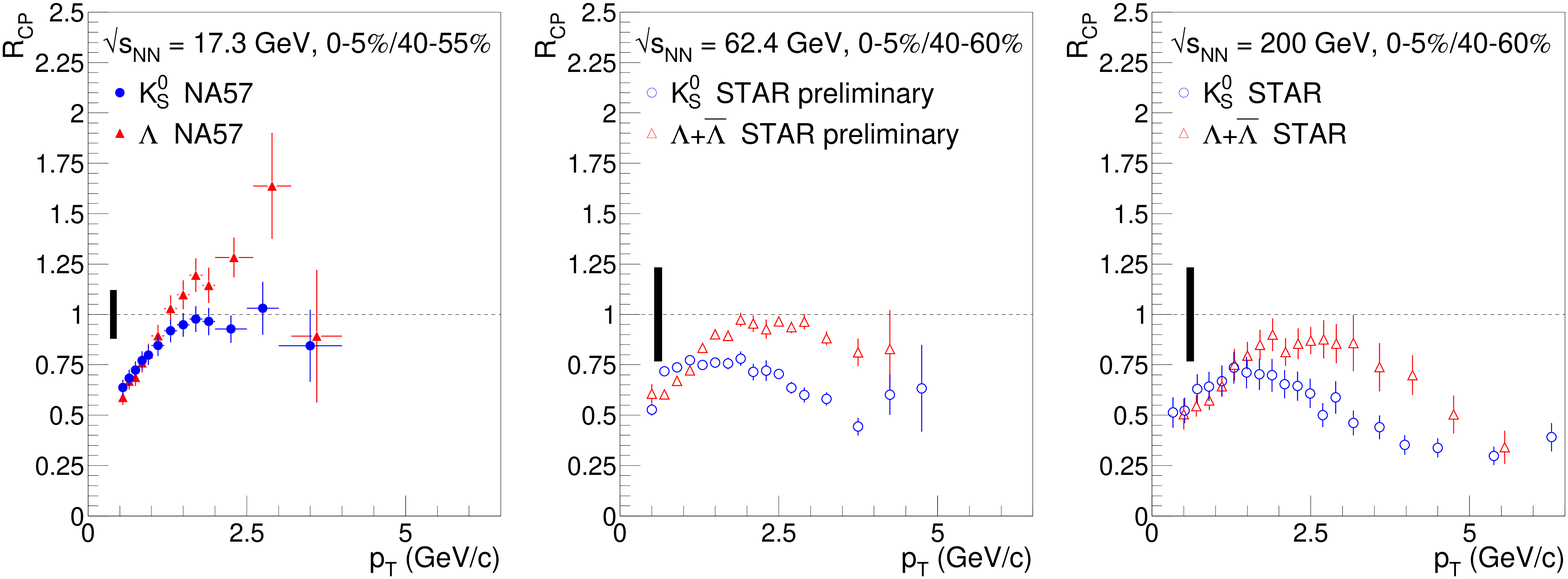}
   \caption{$\RCP(\pt)$ for $\Kzs$ and $\La$ at 
            $\sqrtsNN=17.3~\gev$~\cite{rcppaper} from NA57, and at 
            $\sqrtsNN=62.4~\gev$~\cite{salur} and 
            $200~\gev/c$~\cite{starRcpk0la} from STAR ($\La+\Al$).
            }
   \label{fig:Rcpk0la}
\end{center}
\end{figure}

Figure~\ref{fig:Rcpk0la} shows
the comparison for $\Kzs$ and $\La$ at SPS and RHIC (STAR
data for \mbox{Au--Au} at $\sqrtsNN=62.4~\gev$ (preliminary)~\cite{salur} and 
$200~\gev$~\cite{starRcpk0la}).
In the $\pt$ range covered by our data, up to $4~\gev/c$,
the relative pattern for $\Kzs$ and $\La$ is similar at
the three energies, while absolute values decrease gradually
with increasing energy, from top SPS to top RHIC energy.
The similarity of the $\La$--$\rm K$ pattern
to that observed at RHIC may be taken as an indication for recombination 
effects at SPS energy. This interpretation is also reminiscent of the 
`historic' argument for the strangeness enhancement in a QGP 
scenario, where strange and non-strange quarks would {\it recombine} 
with other quarks in the system to form
strange and multi-strange hyperons. 
We note, however, that the $\La$--$\rm K$ pattern may also be
understood in terms of larger Cronin effect for $\La$ 
with respect to K.

\section{Conclusions}
\label{concl}

In summary, the NA57 experiment has measured:
\begin{Itemize}
\item A hierarchical pattern 
according to strangeness content for hyperon
enhancements ($\pt$-integrated yield per 
participant in \mbox{Pb--Pb} relative to \mbox{p--Be}): 
$E(\La)<E(\Xi)<E(\Omega)$
at 158~$A~\gev/c$ and $E(\La)<E(\Xi)$ at 40~$A~\gev/c$.
This pattern was predicted as a consequence of a phase transition to 
a deconfined quark-gluon plasma~\cite{rafelski}.
\item Strange particles $m_T$ and rapidity distributions
that can be described within a hydrodynamical picture, with a superposition 
of thermal motion with a kinetic freeze-out temperature 
$T\simeq 144~\mev$ and collective flow with similar 
transverse and longitudinal average velocities 
$\av{\beta_\perp}\simeq\av{\beta_{\rm L}}\simeq 0.4$, in the 
centrality range 0--53\% at top SPS energy. 
\item A central-to-peripheral $\RCP$ pattern qualitatively 
similar to that observed at RHIC, although higher in the absolute 
values. In particular, the difference between $\Kzs$ and $\La$ suggests 
a recombination-induced baryon/meson effect also at SPS energy, 
and the comparison of the $\Kzs$ data with theoretical calculations
favours the presence of parton energy loss.
This issue could be clarified
with a systematic analysis of the \mbox{p--Pb}/\mbox{p--Be} 
and \mbox{Pb--Pb}/\mbox{p--Be} 
nuclear modification factors, that would allow to disentangle
Cronin enhancement and energy loss suppression.  
\end{Itemize}


\end{document}